\documentclass[12pt, twoside]{article} 
\usepackage{amssymb}
\usepackage{amscd}
\usepackage{amsmath} 
\usepackage{amsfonts} 
\usepackage{amsthm} 
\usepackage{eucal}
\usepackage{latexsym} 
\theoremstyle{definition} 
\newtheorem{defn}{Definition}[section] 
\theoremstyle{plain} 
\newtheorem{thm}{Theorem}[section] 
\newtheorem{lem}{Lemma}[section]
\newtheorem{cor}{Corollary}[section] 
\theoremstyle{remark} 
\newtheorem{rem}{Remark}[section] 
\newcommand{\field}[1]{\mathbb{#1}}  
\newcommand{\A}{\field{A}}
\newcommand{\F}{\field{F}}  
\newcommand{\R}{\field{R}} 
\newcommand{\HH}{\field{H}} 
\newcommand{\bo}[1]{\boldsymbol{#1}} 
\newcommand{\mf}[1]{\mathsf{#1}} 
\newcommand{\mc}[1]{\mathcal{#1}} 
\begin{document}
\title{Natural FLRW metrics on the Lie group of nonzero quaternions} 
\author{Vladimir Trifonov \\ 
trifonov@member.ams.org} 
\date{}\maketitle 
\begin{abstract} 
It is shown that the Lie group of invertible elements of the quaternion algebra carries a family of natural closed  Friedmann-Lema\^{\i}tre-Robertson-Walker metrics. \end{abstract} 

\section*{Introduction.} The quaternion algebra $\HH$ is one of the most important and well-studies objects in mathematics (e.~g. \cite{Wid02} and references therein) and physics (e.~g. \cite{Adl95} and references therein). It has a natural Hermitian form which induces a Euclidean inner product on its additive vector space $S_{\HH}$. There is also a family of natural Minkowski inner products (signature 2) on $S_{\HH}$, induced by the structure tensor $\bo{\mf{H}}$ of the quaternion algebra. This result was obtained in \cite{Tri95}, where a notion of a natural inner product on a linear algebra over a field $\F$ was introduced. The result came out of a study of relationship between natural metric properties of unital algebras and internal logic of topoi they generate. It was shown in \cite{Tri95} that if the logic of a topos is bivalent Boolean then the generating algebra is isomorphic to the quaternion algebra with a family of Minkowski inner products.  In this note we show that for a unital algebra the inner products can be naturally extended over the Lie group of its invertible elements, producing a family of \emph{principal} \emph{metrics}. In particular, for the quaternion algebra, these metrics are closed Friedmann-Lema\^{\i}tre-Robertson-Walker. These metrics are of interest because they constitute one of the most important classes (as far as \emph{our} universe is concerned) of solutions of Einstein's equations, and there are indications in astrophysics and cosmology that the universe may be spatially closed (\cite{Tr03} and references therein).
\begin{rem} Some of the notations are slightly nonstandard. Small Greek indices, $\alpha, \beta, \gamma$ and small Latin indices $p, q$ \emph{always} run $0$ to $3$ and $1$ to $3$, respectively. Summation is assumed on repeated indices of different levels. We use the $[\begin{smallmatrix} m\\n \end{smallmatrix}]$ device to denote tensor ranks; for example a one-form is a $[\begin{smallmatrix} 0\\1 \end{smallmatrix}]$-tensor. For clarity of the exposition we use $\Box$ at the end of a \emph{Proof}, and each \emph{Remark} ends with the sign appearing at the end of this line. \hfill $\Diamond$ \end{rem}
\begin{defn} An $\F$-\emph{algebra}, $\A$, is an ordered pair $(S_{\A}, \bo{\mf{A}})$, where $S_{\A}$ is a vector space over a field $\F$, and $\bo{\mf{A}}$ is a $[\begin{smallmatrix} 1\\2 \end{smallmatrix}]$-tensor  on $S_{\A}$, called the \emph{structure} \emph{tensor} of $\A$. Each vector $\bo{a}$ of $S_{\A}$ is called an \emph{element} of $\A$, denoted $a \in \A$.  The \emph{dimensionality} of $\A$ is that of $S_{\A}$. \end{defn} 
\begin{rem} This is an unconventional definition of a linear algebra over $\F$. Indeed, the tensor $\bo{\mf{A}}$ induces a binary operation $S_{\A} \times S_{\A} \to S_{\A}$, called the \emph{multiplication} of $\A$: to each pair of vectors $(\bo{a}, \bo{b})$ the tensor $\bo{\mf{A}}$ associates a vector $\bo{ab} : S^*_{\A} \to \F$, such that $(\bo{ab})(\tilde{\bo{\tau}}) =  \bo{\mf{A}}(\tilde{\bo{\tau}}, \bo{a}, \bo{b}), \forall \tilde{\bo{\tau}} \in S^*_{\A}$. An $\F$-algebra with an associative multiplication is called \emph{associative}. An element $\bo{\imath}$, such that $\bo{a\imath} = \bo{\imath a} =\bo{a}, \forall \bo{a} \in \A$ is called an \emph{identity} of $\A$.  \hfill $\Diamond$ \end{rem} 
\begin{defn} For an $\F$-algebra $\A$ and a nonzero one-form $\tilde{\bo{\tau}} \in S^*_{\A}$, a \emph{principal} \emph{inner} \emph{product} is a  $[\begin{smallmatrix} 0\\2 \end{smallmatrix}]$-tensor, $\bo{\mf{A}\left [\tilde{\bo{\tau}}\right ]}$, on $S_{\A}$, assigning to each ordered pair $(\bo{a}, \bo{b})$  a    number $\bo{\mf{A}\left [\tilde{\bo{\tau}}\right ]}(\bo{a}, \bo{b}) := \bo{\mf{A}}(\tilde{\bo{\tau}}, \bo{a}, \bo{b}) \in \F$, just in case it is symmetric, $\bo{\mf{A}\left [\tilde{\bo{\tau}}\right ]}(\bo{a}, \bo{b}) = \bo{\mf{A}\left [\tilde{\bo{\tau}}\right ]}(\bo{b}, \bo{a}), \forall  \bo{a}, \bo{b} \in \A$. \end{defn} 
\begin{rem} In other words, a principal inner product is the contraction of a one-form with the structure tensor. \hfill $\Diamond$ \end{rem} 
\begin{defn} For each $\F$-algebra $\A = (S_{\A}, \bo{\mf{A}})$, an $\F$-algebra $\left [\A\right ] = (S_{\A}$, $\left [\bo{\mf{A}}\right ])$, with the structure tensor defined by  \begin{displaymath} \left [\bo{\mf{A}}\right ](\tilde{\bo{\tau}}, \bo{a}, \bo{b}) := \bo{\mf{A}}(\tilde{\bo{\tau}}, \bo{a}, \bo{b}) - \bo{\mf{A}}(\tilde{\bo{\tau}}, \bo{b}, \bo{a}), \forall \tilde{\bo{\tau}} \in S^*_{\A}, \bo{a}, \bo{b} \in \A , \end{displaymath} 
is called the \emph{commutator} algebra of $\A$. \end{defn}
\begin{defn} A finite dimensional associative $\R$-algebra with an identity is called a \emph{unital} algebra. \end{defn}
\begin{lem} The set $\mc{A}$ of all invertible elements of a unital algebra $\A$ is a Lie group with respect to the multiplication of $\A$, with $\left [\A\right ]$ as its Lie algebra. \end{lem} 
\begin{proof} See, for example, \cite{Pos82} for a proof of this simple lemma. \end{proof} 
\begin{rem} For an $\R$-algebra $\A$, its vector space $S_{\A}$ canonically generates a (linear) manifold $\mc{S}_{\A}$ with the same carrier, with a bijection  $\mc{J} :  \mc{S}_{\A} \to S_{\A}  $. We use the normal ($a, u, ...$) and bold ($\bo{a}, \bo{u}...$) fonts, respectively,  to denote their elements, e.~g., $\mc{J}(a) = \bo{a}$. The tangent space $T_a\mc{S}_{\A}$ is identified with $S_{\A}$ at each point $a \in \mc{S}_{\A}$ via an isomorphism $\mc{J}^*_a : T_a\mc{S}_{\A} \to S_{\A}$ sending a tangent vector to the curve $\mu : \R \to \mc{S}_{\A}, \mu(t) =  a + tu$, at the point $\mu(0) = a \in \mc{S}_{\A}$, to the vector $\bo{u} \in S_{\A}$, with the ``total'' map $\mc{J}^* : T\mc{S}_{\A} \to S_{\A}$. A linear map $\bo{F} : S_{\A} \to S_{\A}$ induces a vector field $\bo{f} : \mc{S}_{\A} \to T\mc{S}_{\A}$ on $\mc{S}_{\A}$, such that the following diagram commutes, 
\begin{equation} \label{VM} \begin{CD} \mc{S}_{\A} @> \bo{f} >> T\mc{S}_{\A} \\ @V \mc{J} VV @VV \mc{J}^* V\\ 
S_\A @> \bo{F} >> S_{\A} \end{CD} \quad . \end{equation} For a unital algebra $\A$ the Lie group $\mc{A}$ is a submanifold of $\mc{S}_{\A}$, with the inclusion map $\bar{\mc{J}} : \mc{A} \to \mc{S}_{\A}$, which is the restiction, to $\mc{A}$, of the identity map. \hfill $\Diamond$ \end{rem}
\begin{rem} \label{FRAMES} For each basis $(\bo{e}_j)$ on the vector space $S_{\A}$ of a unital algebra, there is a natural basis field on $\mc{A}$, namely the basis $(\hat{\bo{e}}_j)$ of left invariant vector fields generated by $(\bo{e}_j)$. We call $(\hat{\bo{e}}_j)$ a \emph{proper} \emph{frame} \emph{generated} \emph{by} $(\bo{e}_j)$. The value, $(\hat{\bo{e}}_j)(a)$, of $(\hat{\bo{e}}_j)$ at $a$ is basis on the tangent space $T_{a}\mc{A}$; it is referred to as a \emph{proper} \emph{basis} (at $a$) generated by $(\bo{e}_j)$. In particular, $(\hat{\bo{e}}_j)(\imath)$, the proper basis at the identity generated by $(\bo{e}_j)$ coincides with $(\bo{e}_j)$. \hfill $\Diamond$ \end{rem}
\begin{defn} For a unital algebra $\A$, let $(\hat{\bo{e}}_j)$ be a  proper frame on $\mc{A}$, generated by a basis $(\bo{e}_j)$ on $S_{\A}$. The \emph{structure} \emph{field} of the Lie group $\mc{A}$ is a tensor field $\bo{\mc{A}}$ on $\mc{A}$, assigning to each point $a \in \mc{A}$ a $[\begin{smallmatrix} 1\\2 \end{smallmatrix}]$-tensor $\bo{\mc{A}(}a\bo{)}$ on $T_a\mc{A}$, with components $\mc{A}^i_{jk}(a) := (\mc{A}(a))^i_{jk}$ in the basis  $(\hat{\bo{e}}_j)(a)$, defined by 
\begin{displaymath} \mc{A}^i_{jk}(a) := \mf{A}^i_{jk} , \quad \forall a \in \mc{A} ,  \end{displaymath} where $\mf{A}^i_{jk}$ are the components of the structure tensor $\bo{\mf{A}}$ in the basis $(\bo{e}_j)$. \end{defn} 
\begin{rem} Intuitively, the structure field is the constant extension of the structure tensor along the left invariant vector fields. \hfill $\Diamond$ \end{rem} 
\begin{defn} For a unital algebra $\A$ and each $a \in \mc{A}$, an $\R$-algebra $\A\{a\} = (S_{\A\{a\}},  \bo{\mf{A}\{}a\bo{\}})$, where $S_{\A\{a\}} := T_{a}\mc{A}$, and $\bo{\mf{A}\{}a\bo{\}} := \bo{\mc{A}(}a\bo{)}$, is called the \emph{tangent} \emph{algebra} of the Lie group $\mc{A}$ at $a$. \end{defn}
\begin{rem} It is easy to see that for each $a \in \mc{A}$, the tangent algebra $\A\{a\}$ is isomorphic to $\A$; in particular, each $\A\{a\}$ is unital. \hfill $\Diamond$ \end{rem} 
\begin{defn} For a unital algebra $\A$ and a twice differentiable real function $\mc{T}$ on the Lie group $\mc{A}$, a \emph{principal} \emph{metric} \emph{on} $\mc{A}$ is a $[\begin{smallmatrix} 0\\2 \end{smallmatrix}]$-tensor field $\bo{\mc{T}}$ on $\mc{A}$, such that that $\bo{\mc{T}}(a) = \bo{\mf{A}\{}a\bo{\}\left [\tilde{a}\right ]}, \forall a \in \mc{A}$, where $\bo{\tilde{a}} := d\mc{T}(a)$ is the value of the gradient of $\mc{T}$ at $a$. \end{defn} 
\begin{rem} In other words, a principal metric is the contraction of a one-form field on $\mc{A}$  with the structure field of $\mc{A}$. For each $a \in \mc{A}$, the value, $\bo{\mc{T}}(a)$, of $\bo{\mc{T}}$ is a principal inner product on the tangent algebra $\A\{a\}$. \hfill $\Diamond$ \end{rem} 
\section{Quaternion algebra.}
\begin{defn} A four dimensional $\R$-algebra, $\HH = (S_{\HH}, \bo{\mf{H}})$, is called a \emph{quaternion} \emph{algebra} (with \emph{quaternions} as its elements), if there is a basis on $S_{\HH}$, in which the components of the structure tensor $\bo{\mf{H}}$ are given by the entries of the following matrices, 
\begin{multline} \label{QST} \mf{H}^0_{\alpha \beta} = 
\begin{pmatrix} 1&0&0&0\\0&-1&0&0\\0&0&-1&0\\0&0&0&-1 \end{pmatrix},\ 
\mf{H}^1_{\alpha \beta} = \begin{pmatrix} 0&1&0&0\\1&0&0&0\\0&0&0&1\\0&0&-
1&0 \end{pmatrix}, \\ \mf{H}^2_{\alpha \beta} = \begin{pmatrix} 
0&0&1&0\\0&0&0&-1\\1&0&0&0\\0&1&0&0 \end{pmatrix},\ \mf{H}^3_{\alpha 
\beta} = \begin{pmatrix} 0&0&0&1\\0&0&1&0\\0&-1&0&0\\1&0&0&0 
\end{pmatrix}. \end{multline} We refer to such a basis as \emph{canonical}. \end{defn} 
\begin{rem} The vectors of the canonical basis are denoted $\bo{1}$, $\bo{i}$, $\bo{j}$, $\bo{k}$. A quaternion algebra is unital, with the first vector  of the canonical basis, $\bo{1}$, as its identity. Since $(\bo{1}$, $\bo{i}$, $\bo{j}$, $\bo{k})$ is a basis on a real vector space, any quaternion $a$ can be presented as $a^0\bo{1} + a^1\bo{i} + a^2\bo{j} + a^3\bo{k}, a^{\beta} \in \R$. A quaternion $\bar{a} = a^0\bo{1} - a^1\bo{i} - a^2\bo{j} - a^3\bo{k}$ is called \emph{conjugate} to $a$. We refer to $a^0$ and $a^p\bo{i}_p$ as the \emph{real} and  \emph{imaginary} \emph{part} of $a$, respectively. Quaternions of the form $a^0\bo{1}$ are in one-to-one correspondence with real numbers, which is often denoted, with certain notational abuse, as $\R \subset \HH$. \hfill $\Diamond$ \end{rem}
\begin{rem} \label{SO(3)} A linear transformation $S_{\HH} \to S_{\HH}$ with the following components in the canonical basis, 
\begin{displaymath} \begin{pmatrix} 1 & 0 \\ 0& \bo{\mf{B}} \end{pmatrix}, \bo{\mf{B}} \in SO(3), \end{displaymath} 
takes $(\bo{1}$, $\bo{i}$, $\bo{j}$, $\bo{k})$ to a basis $(\bo{i}_{\beta})$ in which the components \eqref{QST} of the structure tensor will \emph{not} change, and neither will the multiplicative behavior of vectors of $(\bo{i}_{\beta})$. Thus, we have a class of canonical bases parameterized by elements of $SO(3)$. \hfill $\Diamond$ \end{rem}
\begin{lem} Every principal inner product on $\HH$ is Minkowski. \end{lem} 
\begin{proof} For the quaternion algebra the components of the structure tensor $\bo{\mf{H}}$ in a canonical basis are given by \eqref{QST}. 
A one-form $\tilde{\bo{\tau}}$ on $S_{\HH}$ with components $\tilde{\tau}_{\beta}$ in (the dual of) a canonical basis $(\bo{i}_{\beta})$ contracts with the structure tensor into a $[\begin{smallmatrix} 0\\2 \end{smallmatrix}]$-tensor on $S_{\HH}$ with the following components in the basis $(\bo{i}_{\beta})$: 
\begin{displaymath} \begin{pmatrix} 
\tilde{\tau}_0& \tilde{\tau}_1& \tilde{\tau}_2& \tilde{\tau}_3\\ 
\tilde{\tau}_1&-\tilde{\tau}_0& \tilde{\tau}_3&-\tilde{\tau}_2\\ 
\tilde{\tau}_2&-\tilde{\tau}_3&-\tilde{\tau}_0& \tilde{\tau}_1\\ 
\tilde{\tau}_3& \tilde{\tau}_2&-\tilde{\tau}_1&-\tilde{\tau}_0 \end{pmatrix}. \end{displaymath} 
The only way to make this symmetric is to put $\tilde{\tau}_1=-\tilde{\tau}_1$, $\tilde{\tau}_2=-\tilde{\tau}_2$,  $\tilde{\tau}_3=-\tilde{\tau}_3$, which yields $\tilde{\tau}_1=\tilde{\tau}_2=\tilde{\tau}_3=0$: 
\begin{equation} \label{PSP} (\mf{H}\left [\tilde{\bo{\tau}}\right ])_{\alpha \beta} = \begin{pmatrix} 
\tilde{\tau}_0& 0& 0& 0\\ 
0&-\tilde{\tau}_0& 0&0\\ 
0&0&-\tilde{\tau}_0& 0\\ 
0& 0&0&-\tilde{\tau}_0 \end{pmatrix}. \end{equation} . \end{proof}
\section{Natural structures on $\mc{H}$.} There is a class of canonical bases   
on $S_{\HH}$ (see Remark \ref{SO(3)}) whose members differ from one another by a rotation in the hyperplane of pure 
imaginary quaternions. Each canonical basis $(\bo{i}_{\beta})$ on $S_{\HH}$ induces a \emph{canonical}  
coordinate system $(w$, $x$, $y$, $z)$ on the linear manifold $\mc{S}_{\HH}$ canonically generated by $S_{\HH}$, and 
therefore also on its submanifold $\mc{H}$ of nonzero quaternions: a point $a \in \mc{H}$ such that $\mc{J}(a) = \bo{a} = a^{\beta}\bo{i}_{\beta}$ 
is assigned coordinates $(w = a^0$, $x = a^1$, $y = a^2$, $z = a^3)$. This coordinate 
system covers both $\mc{S}_{\HH}$ and $\mc{H}$ with a single patch. Since $\bo{0} 
\notin \mc{H}$, at least one of the coordinates is always nonzero for any point $a \in 
\mc{H}$. For a differentiable function $R : \R \to \R \setminus \{0\}$ there is a system of natural spherical coordinates  $(\eta$, $\chi$, $\theta$, $\varphi)$ on $\mc{H}$, related to the canonical coordinates by 
\begin{multline*} 
w = R(\eta)\cos\chi, \quad x = R(\eta)\sin\chi\sin\theta\cos\varphi, \\
y = R(\eta)\sin\chi\sin\theta\sin\varphi, \quad z = R(\eta)\sin\chi\cos\theta . 
\end{multline*} 
Each canonical basis $(\bo{i}_{\beta})$ 
can be considered a basis on the vector space of the Lie algebra of $\mc{H}$, i.~e., 
the tangent space $T_{\bo{1}}\mc{H} \cong S_{\HH}$ to $\mc{H}$ at the point $(1$, $0$, 
$0$, $0)$, the identity of the group $\mc{H}$. There are several natural basis fields on 
$\mc{H}$ induced by each basis $(\bo{i}_{\beta})$. First of all, we have  the proper frame  $(\hat{\bo{\imath}}_{\beta})$, of left invariant vector fields on $\mc{H}$ (see Remark \ref{FRAMES}), which is a \emph{noncoordinate} basis field. There are also two \emph{coordinate} 
basis fields, the \emph{canonical} \emph{frame}, $(\partial_w$, $\partial_x$, $\partial_y$, $\partial_z)$ and the corresponding \emph{spherical} \emph{frame}  $(\partial^R_{\eta}$, $\partial^R_{\chi}$, $\partial^R_{\theta}$, $\partial^R_{\varphi})$.  A left invariant vector field $\hat{\bo{u}}$ on $\mc{H}$, generated by a vector $\bo{u} \in S_{\HH}$ with components $(u^{\beta})$ in a canonical basis, associates to each point $a \in \mc{H}$ with coordinates $(w$, $x$, $y$, $z)$ a vector $\hat{\bo{u}}(a) \in T_{a}\mc{H}$ with the components $\hat{u}^{\beta}(a) = (\bo{a}\bo{u})^{\beta}$ in the basis $(\partial_w$, $\partial_x$, $\partial_y$, $\partial_z)(a)$ on $T_{a}\mc{H}$: 
\begin{multline} \label{LVFIELDS} 
\hat{u}^0(a) = wu^0 - xu^1 - yu^2 - zu^3 , \quad \hat{u}^1(a) = wu^1 + xu^0 + yu^3 - zu^2 , \\ 
\hat{u}^2(a) = wu^2 - xu^3 + yu^0 + zu^1 , \quad \hat{u}^3(a) = wu^3 + xu^2 - yu^1 + zu^0 . 
\end{multline} 
The system \eqref{LVFIELDS} contains sufficient information to compute transformation between the frames. 
For example, the transformation between the spherical and proper frames is given by 
\begin{displaymath} \begin{pmatrix} 
R/\dot{R} & 0 & 0 & 0\\ 
0 & \sin{\theta}\cos{\varphi} & \sin{\theta}\sin{\varphi} & \cos{\theta} \\ 
0 & \frac{\cos{\chi}\cos{\theta}\cos{\varphi}+\sin{\chi}\sin{\varphi}}{\sin{\chi}} & 
\frac{\cos{\chi}\cos{\theta}\sin{\varphi}+\sin{\chi}\cos{\varphi}}{\sin{\chi}} & 
\frac{\cos{\chi}\sin{\theta}}{\sin{\chi}}\\ 
0 & \frac{\sin{\chi}\cos{\theta}\cos{\varphi}-\cos{\chi}\sin{\varphi}}{\sin{\chi}\sin{\theta}} & 
\frac{\sin{\chi}\cos{\theta}\sin{\varphi}+\cos{\chi}\cos{\varphi}}{\sin{\chi}\sin{\theta}} & -1 
\end{pmatrix}, \end{displaymath} where $\dot{R} := \frac{dR}{d{\eta}} : \R \to \R \setminus \{0\}$. 
\begin{defn} A Lorentzian metric on a four dimensional manifold is called \emph{closed} \emph{FLRW} (Friedmann-Lema\^{\i}tre-Robertson-Walker) if there is a coordinate system $(x^{\beta})$, such that in the corresponding coordinate frame the components of the metric are given by the entries of the following matrix: \begin{displaymath} 
\genfrac{}{}{0pt}{3}{+}{-} \begin{pmatrix} 
1&0&0&0\\ 0&-\mf{a}^2&0&0\\ 0&0&-\mf{a}^2\sin^2(x^1) &0\\ 
0&0&0&-\mf{a}^2\sin^2(x^1)\sin^2(x^2) \end{pmatrix}, \end{displaymath} where $\mf{a} : \R \to \R$, referred to as the \emph{scale} \emph{factor}, is a function of $x^0$ only. \end{defn}
\section{Principal metrics on $\mc{H}$.}
\begin{thm} Every principal metric on $\mc{H}$ is closed FLRW. \end{thm} 
\begin{proof} Let $\tilde{\bo{\tau}}$ and $(\bo{i}_{\beta})$ be a one-form and a canonical basis on $S_{\HH}$, respectively. For each point $a \in \mc{H}$ the $\R$-algebra $\HH(a)$ is the tangent algebra, at $a$, of the Lie group $\mc{H}$. For each $a \in \mc{H}$ the components of the structure tensor  $\bo{\mf{H}\{}a\bo{\}}$ and a principal inner product, $\bo{\mf{H}\{}a\bo{\}\left [\tilde{\tau}\right ]}$, of $\HH(a)$ in the basis  $(\bo{\hat{\imath}}_{\beta})(a)$ are given by \eqref{QST} and  \eqref{PSP}, respectively. Therefore, the components of a principal metric, $\bo{\mc{T}}$, in the proper frame $(\bo{\hat{\imath}}_{\beta})$ must have the form 
\begin{equation} \begin{pmatrix} 
\tilde{\tau}& 0& 0& 0\\ 
0&-\tilde{\tau}& 0&0\\ 
0&0&-\tilde{\tau}& 0\\ 
0& 0&0&-\tilde{\tau} \end{pmatrix}, \end{equation} 
for some function $\tilde{\tau} : \mc{H} \to \R \setminus \{0\}$. In other words, any principal metric on $\mc{H}$ is obtained by contraction of a one-form field $\tilde{\bo{\tau}}$, whose components in $(\bo{\hat{\imath}}_{\beta})$ are $(\tilde{\tau}, 0, 0, 0)$, with the structure field $\bo{\mc{H}}$. This one-form is exact, i.~e., there exists a twice differentiable function $\mc{T}$, such that $d\mc{T} = \tilde{\bo{\tau}}$. In the spherical frame with $R(\eta) = \exp(\eta)$ the components of $\tilde{\bo{\tau}}$ are also $(\tilde{\tau}$, $0$, $0$, $0)$ , and,   
\begin{equation} \label{TIME} d\mc{T}_0 = \frac{\partial \mc{T}}{\partial \eta} = \tilde{\tau}, \quad d\mc{T}_1 = \frac{\partial \mc{T}}{\partial \chi} = d\mc{T}_2 = \frac{\partial \mc{T}}{\partial \theta} = d\mc{T}_3 = \frac{\partial \mc{T}}{\partial \varphi} = 0 . \end{equation} 
It follows from \eqref{TIME} that both $\mc{T}$ and $\tilde{\tau}$ depend on $\eta$ only. Since $\frac{\partial \mc{T}}{\partial \eta}$ is differentiable, $\tilde{\tau}$ must be at least continuous. Since $\tilde{\tau}(\eta) \neq 0, \forall \eta \in \R$, $\tilde{\tau}$ cannot change sign. Computing the components of the principal metric $\bo{\mc{T}}$ in the spherical frame we get
\begin{displaymath} \mc{T}_{\alpha \beta} = \begin{pmatrix} 
\tilde{\tau}(\eta)(\frac{\dot{R}}{R})^2&0&0&0\\0&-\tilde{\tau}(\eta)&0&0\\ 
0&0&-\tilde{\tau}(\eta){\sin^2\chi} &0\\ 
0&0&0&-\tilde{\tau}(\eta){\sin^2\chi} {\sin^2\theta} \end{pmatrix}.  \end{displaymath}  
If $\tilde{\tau}(\eta) > 0$, we take $R(\eta)$ such that $\tilde{\tau}(\eta)(\frac{\dot{R}}{R})^2 = 
1$, which yields \begin{equation} \label{+R} R(\eta) = 
exp{\int\frac{d\eta}{\genfrac{}{}{0pt}{3}{+}{-}\sqrt{\tilde{\tau}(\eta)}}} \quad . \end{equation} 
In other words, with $R(\eta)$ satisfying \eqref{+R}, the components of the principal metric in the spherical frame are 
\begin{displaymath} 
\mc{T}_{\alpha \beta} = \begin{pmatrix} 
1&0&0&0\\ 0&-\mf{a}^2&0&0\\ 0&0&-\mf{a}^2{\sin^2\chi} &0\\ 
0&0&0&-\mf{a}^2{\sin^2\chi} {\sin^2\theta} \end{pmatrix}, \end{displaymath} 
where the scale factor $\mf{a}(\eta) := {\sqrt{\tilde{\tau}(\eta)}}$. \par 
If $\tilde{\tau}(\eta) < 0$, similar considerations show that the metric is also closed FLRW with the scale factor $\mf{a}(\eta) := {\sqrt{-\tilde{\tau}(\eta)}}$. \end{proof} 
\begin{cor} $\mc{T}$ is a monotonous function of $\eta$ for each principal metric $\bo{\mc{T}}$ of $\mc{H}$. \end{cor} 
Thus the natural geometry of the Lie group of nonzero quaternions $\mc{H}$ is defined by a family of closed  Friedmann-Lema\^{\i}tre-Robertson-Walker metrics. 

\end{document}